\documentclass{nature}
\usepackage[symbol]{footmisc}
\usepackage{graphicx}
\usepackage{amsmath}
\usepackage[usenames]{color}
\usepackage{url}
\usepackage{lscape}
\usepackage{booktabs,multirow}
\title{Are machine learning technologies ready to be used for humanitarian work and development?}

\author{Vedran Sekara$^{1,2,*}$, M\'arton Karsai$^{3,4}$, Esteban Moro$^{5,6}$, Dohyung Kim$^1$, Enrique Delamonica$^1$, Manuel Cebrian$^{7}$, Miguel Luengo-Oroz$^8$, Rebeca Moreno Jim\'enez$^9$ \& Manuel Garcia-Herranz$^{1,*}$}

\begin{document}

\maketitle
\linespread{1}
\begin{affiliations}
 \item UNICEF, New York, USA
 \item IT University of Copenhagen, Denmark
 \item Central European University, Vienna, Austria 
 \item R\'enyi Institute of Mathematics, Budapest, Austria
 \item Connection Science, Massachusetts Institute of Technology, Cambridge, MA, USA
 \item Department of Mathematics \& GISC, Universidad Carlos III de Madrid, Spain
 \item Department of Statistics, Universidad Carlos III de Madrid, Spain
 \item United Nations Global Pulse, New York, USA
 \item UNHCR, Geneva, Switzerland
 
 \item[*] Correspondence should be addressed to these authors
\end{affiliations}

\begin{abstract}
Novel digital data sources and tools like machine learning (ML) and artificial intelligence (AI) have the potential to revolutionize data about development and can contribute to monitoring and mitigating humanitarian problems.
The potential of applying novel technologies to solving some of humanity's most pressing issues has garnered interest outside the traditional disciplines studying and working on international development.
Today, scientific communities in fields like Computational Social Science, Network Science, Complex Systems, Human Computer Interaction, Machine Learning, and the broader AI field are increasingly starting to pay attention to these pressing issues. 
However, are sophisticated data driven tools ready to be used for solving real-world problems with imperfect data and of staggering complexity?
We outline the current state-of-the-art and identify barriers, which need to be surmounted in order for data-driven technologies to become useful in humanitarian and development contexts.
We argue that, without organized and purposeful efforts, these new technologies risk at best falling short of promised goals, at worst they can increase inequality, amplify discrimination, and infringe upon human rights.
\end{abstract}

\section*{New Tools and Datasets}
Data is critical for humanitarian and development work.
Accurate and updated estimates of population demographics are vital in order to understand and respond to social and economic inequalities\cite{ocha2019data}, and to move from reactive to proactive interventions that mitigate the impact of crises before they happen\cite{unocha-anticipatory}.
As such, whether using global estimates of poverty to advocate for efforts or design policies to eliminate it, or using malnutrition data in the midst of a conflict to allocate resources to where they are most needed, data is at the core of the organizations that work to meet the 2030 Agenda for Sustainable Development\cite{united2015transforming}.
It can, however, be hard to obtain accurate and timely data.
In many parts of the world traditional household surveys are the main, and often only, method for demographic data collection.
Surveys provide rich and irreplaceable data, but they can be expensive and time-consuming.
As such, there is a growing focus on leveraging different big digital datasets and new tools like AI and ML to complement household surveys.
This is particularity important in rapidly changing contexts (e.g. humanitarian crises or pandemics) as data and information can be retrieved and analyzed in fast and relatively inexpensive ways.

Unfortunately there are no clear definitions of AI. 
In general terms they refer to systems, which sift through data, recognize patterns, and possibly make decisions based on their
discoveries.
This covers the full spectrum of models, from relatively simple statistical models (e.g. linear regression and decision trees) to more sophisticated, but still explainable mathematical models, to black-box like neural network and deep learning approaches.

Technologies, like mobile phones, are starting to have significant global coverage.
Today there are 107 mobile-cellular subscriptions per 100 inhabitants worldwide\cite{international2018measuring} (see Fig. 1A), 95\% of the world's population is covered by at least a 2G connection\cite{gsma2017unlocking}, and mobile broadband adoption has grown 14-fold from 5 in 2008 to almost 70 subscriptions per 100 inhabitants in 2018\cite{international2018measuring}\footnote{Nonetheless, this does not mean phones and access to broadband are spread evenly across, and within, countries. Often new technologies reproduce and perpetuate existing inequalities.}.
As a consequence, mobile phones and the vast amount of data they produce (incl. social media data) can potentially be applied to tackle humanitarian problems in places, which previously were deemed hard to reach.
In addition, high-resolution satellite images are becoming more readily available with commercial vendors delivering 30cm resolution imagery, while the European Space Agency and NASA (National Aeronautics and Space Administration) are open-sourcing a wide range of remote sensed datasets (see Fig. 1B).
These images analyzed with the support of AI tools can aid humanitarian response efforts, from mapping refugee shelters\cite{quinn2018humanitarian} to quantifying the extent of flooded areas\cite{brivio2002integration}.

The private sector has also started to play a larger role as data providers, and looking to develop new methodologies and business models that can help tackle societal problems under the umbrella of data for public good\cite{eu2020towards}.
Examples of data collaborations between the private sector and academia include the Data for Development (D4D)\cite{d4d} and Data for Refugees (D4R) challenges\cite{salah2019guide}, where telephone companies shared anonymized and aggregated call detail records (CDR) with the goal of contributing to socio-economic development and well-being for the most marginalized populations.
Similarly, GSMA, the industry organization that represents global mobile communications companies, through their \textit{Mobile for Humanitarian Innovation} initiative has opened up for both funding and data access\cite{gsma}.
Nonetheless, these initiatives have also raised serious privacy concerns\cite{maxmen2019surveillance}.
During the COVID-19 pandemic, many other companies opened up their data, examples include Apple, Cuebiq, Facebook, and Google\footnote{Ordered alphabetically, Apple's Mobility Trends Reports \url{https://covid19.apple.com/mobility}, Cuebiq's Data for Good initative \url{https://www.cuebiq.com/about/data-for-good/}, Facebook's Data for Good platform \url{https://dataforgood.fb.com/}, and Google's COVID-19 Community Mobility Reports \url{https://www.google.com/covid19/mobility/}}.
In alignment with this, UN agencies have started to build capacity to collaborate more meaningfully with the private sector on a data level\cite{magicbox,d4c,wb-bigdata}.

These, and similar efforts, have resulted in a plethora of scientific studies (see Fig. 1C) , which focus on combining novel digital data sources (including mobile phone and social media data) with powerful tools from computer science, mathematics, and physics to estimate developmental indicators ranging from: socio-economic status\cite{blumenstock2015predicting,pokhriyal2017combining,abitbol2018location,espin2023interpreting}, illiteracy\cite{schmid2017constructing}, unemployment\cite{llorente2015social,almaatouq2016mobile}, gender inequality\cite{garcia2018analyzing,gauvin2020gender}, segregation\cite{athey2021,moro2021}, and population statistics\cite{deville2014dynamic}.
Similarly, these datasets and mathematical techniques can be used to achieve the Sustainable Development Goals~\cite{vinuesa2020role,porciello2020accelerating,cowls2021definition}.
For instance, data from search engines has been used to understand the determinants of suicides\cite{adler2019search} and chronic health conditions\cite{oladeji2021monitoring}, and AI analyses of satellite images have been applied for similar endeavours, from estimating monetary poverty\cite{jean2016combining,engstrom2017poverty}, to measuring crop type\cite{nowakowski2021crop}, and crop productivity\cite{burke2017satellite}.

Taken together, these novel data sources and tools have the potential to transform international development and humanitarian work.
However, as we argue below they are not yet ready to be rolled out on a large scale.

\begin{figure}[!htbp]
\centering
\includegraphics[width=0.5\linewidth]{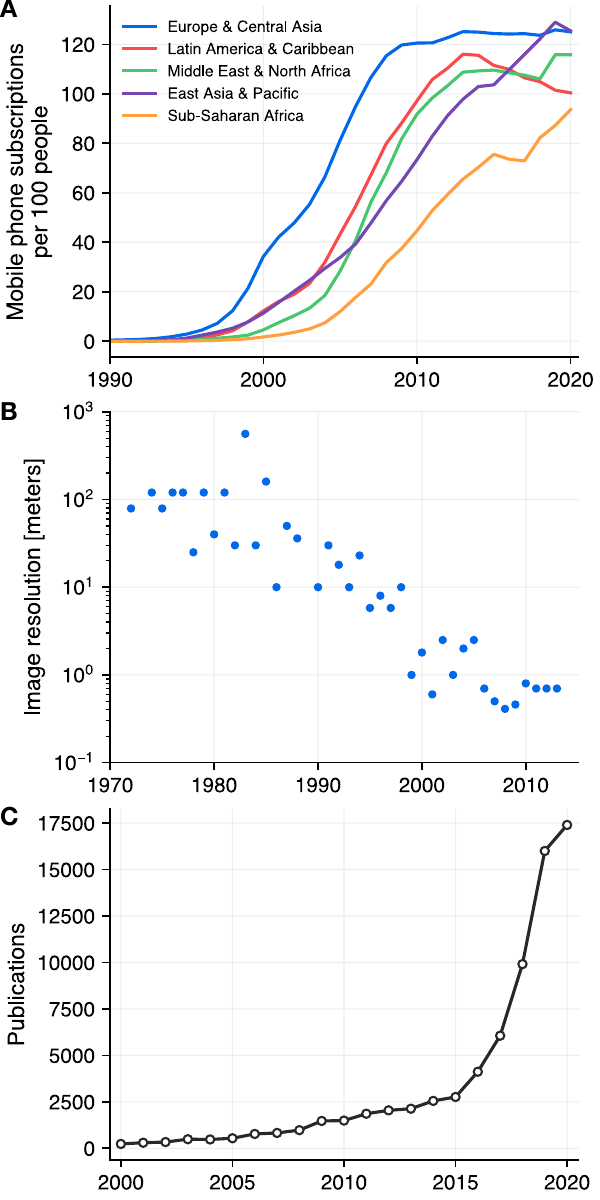}
\caption{\textbf{New technologies and tools.} \textbf{A}, The percentage of people owning a phone is rapidly rising worldwide\cite{international2018measuring}. \textbf{B}, Satellites are increasingly able to capture our world in greater detail. Data shows the maximal aperture resolution per year for all civilian and commercially launched satellites\cite{belward2015launched}. \textbf{C}, The explosive growth of interest in the humanitarian field as reflected in number of papers per year; data from Google Scholar retrieved using the query ("artificial intelligence" OR "machine learning") AND ("sustainable development" OR "humanitarian action" OR "sdgs" OR "millennium development")}
\label{fig:1}
\end{figure}

\section*{Barriers to Useful Tools}
Data does not translate easily into knowledge, it requires careful collection, curation, and aggregation to become informative.
When it comes to AI and ML, there are currently many open issues and challenges.
For instance, the carbon footprint of AI is large, in some cases training one model can emit as much CO$_2$ as 57 average humans emit in a year\cite{strubell2019energy}.
Access to the computational resources needed to power AI and ML technologies is not equally distributed, which can lead to power being concentrated in the hands of a few countries and companies\cite{crawford2021atlas}. 
The use of new AI technologies, such as generative AI threatens to undermine our societies and erode trust in democracies.
Further, when it comes to the application of AI and ML technologies no globally accepted set of AI ethics exist\cite{jobin2019global}, and many have argued that ethical principles alone will not guarantee ethical applications~\cite{mittelstadt2019principles}.
The issues are many, we have identified three which we believe form the biggest obstacles for using AI and ML technologies in humanitarian and development contexts.
We focus on these, not because the others are less important, but because we believe these are often overlooked, and need to be addressed in order for data-driven technologies to become useful in humanitarian and development contexts.

\noindent \textbf{1. The ecosystem of data for development is not yet machine friendly.}
Building any kind of ML or AI model requires access to high-quality data. 
Yet getting access to such datasets can be a hard and laborious process.
Some global survey data are accessible through platforms such as the MICS\footnote{\url{https://mics.unicef.org/surveys/}} (Multiple Indicator Cluster Survey) and DHS\footnote{\url{https://dhsprogram.com/data/}} (Demographic and Health Survey) sites.
However, these datasets are collected, curated, organized, and maintained for the purpose of informing decision makers, not for training algorithms. 
In addition, the vast majority of these datasets are spread across a multitude of national, non-governmental, and international organizations, where data are often locked up in non-machine-readable or proprietary formats and subject to complex or opaque licensing and use regimes that make them difficult to use for ML or automated processes.

Another obstacle arises from non-standardized metrics, where it is entirely possible that two surveys (even within the same country) use different definitions of the same indicator.
For instance, poverty is in certain cases measured by the difference in income or consumption to the average level, regardless of whether this level is sufficient to maintain a decent standard of living, while in other cases poverty is based on being able to afford a minimum amount of various food types and other necessities like shelter.
Different definitions exist because poverty is a complex multifaceted problem, but this breaks with the current thinking in the computational field and makes it hard to benchmark and compare AI models over time and across countries.

Additional issues arise from insufficient or missing geographic information.
Most surveys are georeferenced, but that does not necessarily mean that each individual data-point is labeled with a GPS coordinate. 
This is a reasonable choice if data is only used for decision making, but a severe barrier for using data to train statistical algorithms.
Rather, data are stratified into clusters according to various administrative boundaries such as municipalities, health zones, or census tracts.
For instance, developmental estimates are often reported on a regional level\cite{khan2019multiple}.
Yet, only a minority of surveys are accompanied by a so-called shapefile, which contains information about the geographical boundaries of those administrative regions.
This is an issue as administrative boundaries are not fixed in time, and many times across data collection efforts, they are often redrawn in response to changing populations, conflicts, contested borders, etc.
As such, researchers and practitioners are often left on their own to infer which shapefiles were originally used, or to re-create their own files from old maps.
Lacking information on where development indicators were originally collected makes it challenging to link survey data to insights from new digital data sources.

All things considered, by not making data for development ready to be utilized by new ML and AI tools, we risk impeding the application of these new methodologies towards addressing complex societal issues.
For instance, a lack of machine friendly data might divert the focus of scientific studies to regions, groups, or issues for which machine readable data already exists.
Coordinated efforts to develop standardized open data formats and repositories that contain machine readable development data could save practitioners and researchers countless hours, which instead could be spent on addressing the problems of marginalized communities.
Unfortunately, the AI and ML communities do not yet have standardized processes in place for documenting datasets, but the recently proposed \textit{datasheets for datasets} is a good place to start\cite{gebru2018datasheets}, along with adapting already existing standards that have been developed by national statistics institutes.
Platforms like the Humanitarian Data Exchange\cite{hdx}, built by the United Nations Office for the Coordination of Humanitarian Affairs, are a good first effort for opening data about humanitarian problems, but more efforts are needed if these platforms are to become useful for AI practitioners.

\noindent \textbf{2. Lack of validation, transferability, and generalization of machine models.}
Using digital data to produce population demographics is a relatively new endeavour. 
While the field has produced some exciting results, for instance, fine grained wealth distribution maps have be generated for a large number of countries\cite{blumenstock2015predicting,jean2016combining,pokhriyal2017combining,lee2020high,chi2021micro}, little is known about the shortcomings of these new approaches.
By contrast, household surveys have been used for decades and their limitations are well understood and documented.
With digital data it is unclear whether a model trained on satellite imagery from one period will work on images captured during a different season or if a model based on mobile phone data (or social media data) will work on behavioral traces collected from a different month.

Adding to this concern, it is unclear how these new methodologies transfer across a multitude of different countries, cultures, and contexts.
Some research groups have tried to replicate and benchmark published methodologies with varying success.
For example, Fernando et al.\cite{fernando2018predicting} found that statistical patterns associated with socio-economic characteristics fundamentally differ for western and northern parts of Sri Lanka, Blumenstock\cite{blumenstock2018estimating} demonstrated that models trained on mobile phone data from Rwanda cannot be applied to the context of Afghanistan, while Tingzon et al.\cite{Tingzon2019Mapping} successfully replicated an AI methodology (originally piloted in five African countries) for the Philippines.
Our own experiences in transferring algorithms across countries suggest that relationships between behavioral patterns extracted from digital data and development indicators can greatly differ, even between countries within same geographic regions (see Figure 2A).
For instance, we find the correlation between the diversity of mobility data (shown to be highly indicative of economic development\cite{pappalardo2016analytical}) and the human development index\cite{hdi} to be different between countries, even within the same region (e.g. Costa Rica and Colombia).

The issue of whether models will generalize over longer timescales is also of concern\cite{lazer2014parable}.
As human behavior changes and evolves over time, it is uncertain how robust the inferred statistical relations between demographic variables and digital traces will be.
For instance, a decade ago Eagle et al.\cite{eagle2010network} demonstrated that features extracted from landline communications (in combination with mobile phone interactions) were good indicators of economic development.
But a decade ago smartphones and online social media were just starting to appear, and the subsequent widespread adoption of these technologies raises the question of whether the same model would work equally well today.
All data-driven methods are strongly influenced by technological drift and these effects need to be understood and accounted for if we are to use these technologies for long-term policy making.
However, model generalization is not only affected by long-term shifts in technology usage.
Our experiences in using aggregate behaviors extracted from mobile phone traces show that ML models can be brittle even on relatively short timescales, with model performance dramatically degrading from one month to the next (see Figure 2B).
Regrettably, research centered on this topic is rare in the academic literature, instead the focus has been one-off studies that achieve spectacular results, rather than monitoring and evaluating the robustness of algorithms. 

\begin{figure}[!htbp]
\centering
\includegraphics[width=\linewidth]{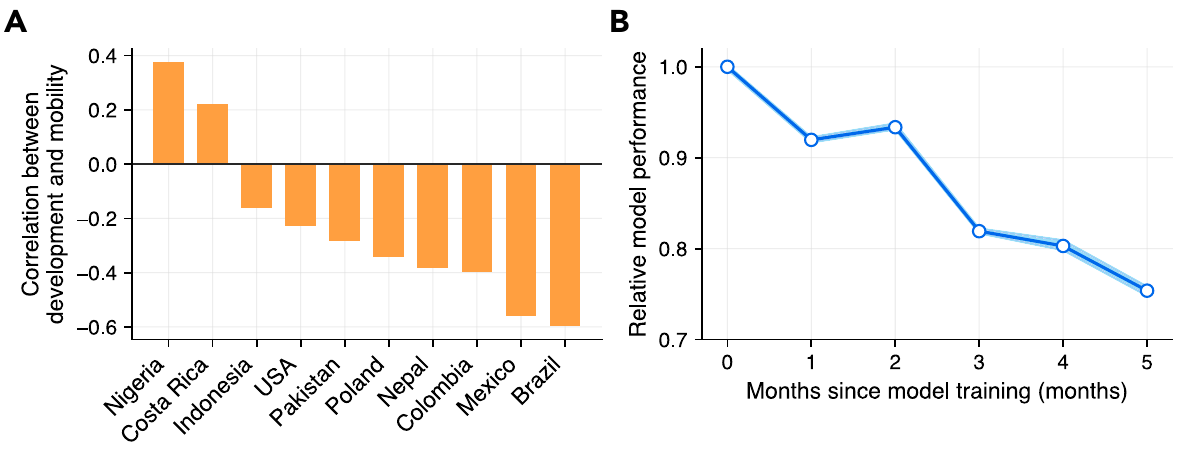}
\caption{\textbf{Common pitfalls of data-driven tools.} \textbf{A}, Models are not easily transferred between contexts, illustrated here by the correlation between the human development index (HDI) and observed mobility from social media. The correlation is calculated between HDI and the entropy $S$ of mobility data, where $S_i=-\sum_{j} p_{ij} \log{p_{ij}}$ and $p_{ij}$ is the probability of observing trips between regions $i$ and $j$. 
\textbf{B}, When models are validated on digital data from later periods, in this case mobile phone data from Iraq, we observe that model performance drastically decays over time. In this case we train a ML model to estimate poverty from mobile phone data using one month to train the model and evaluate how the model performs on data collected 1-5 months later. Model performance is measured using a cross-validated correlation coefficient. Here we show the relative model performance, which is rescaled with respect to the initial accuracy, to quantify how well the model performs for later months. Light blue area denotes the variation across multiple models trained using different cross-validation sets.}
\label{fig:2}
\end{figure}

\noindent\textbf{3. We need to look beyond single metrics.}
While digital technologies might provide advantages to society, automated systems bring learned biases into decision making, leading to new kinds of vulnerabilities\cite{crawford2016there}.
Well known examples include gender, age, racial, class, ability, and wealth biases\cite{o2016weapons,mehrabi2021survey}.
Although researchers have identified some of these vulnerabilities, quantifying others can be difficult due to the complexity and opaqueness of automated algorithmic systems\cite{raji2020closing}.
To address these issues it is important that we look beyond single metrics such as averages, R-squares, and correlation coefficients and start to unpack algorithmic impact across an intersection of inequalities\cite{cho2013toward}.
For instance, reporting an $80\%$ accuracy of an algorithm is not enough as aggregate metrics can hide a lot of nuance.
This is also called the tyranny of averages\cite{annan2018data}. 
Instead we need to disaggregate algorithmic performance, for instance, into how well algorithms works for rural vs urban areas, how well they perform for poor vs wealthy regions, or along other delineations, which might be mis- or underrepresented.
Ultimately, we want to avoid situations where an algorithm may have
an overall accuracy of 80\%, but only work 10\% of the time in poor regions.
As such, we need to ensure algorithmic equity---the (un)equal distribution of algorithmic accuracy across different groups---is measured and reported in future studies.

For humanitarian and development work the goal is not to build an AI tool that can achieve a perfect prediction, rather it is to gain a deeper understanding of gaps and structural inequalities and how to fix them.
In these situations, outliers and incorrect predictions often turn out to be vital discoveries---critical to not leaving anyone behind---which traditional metrics might not identify. 

It is important to be critical of the data used to develop models and draw conclusions from, as it might suffer from strong observational and selection biases.
The excitement of the first years of the \textit{Big Data Revolution}, where focus was on volume and velocity of datasets, needs to change; instead of being obsessed by the number of rows a dataset contains or how many terabytes it takes up, we need to dedicate far more efforts to understand which demographics are left out.
For instance, digital datasets are limited to groups which: own a mobile phone, use social media, have a specific smartphone app installed, or live in a region densely photographed by satellites.
Marginalized communities lacking access to new technologies, or refusing to adopt them\cite{rosenberg2021other}, will be unobservable in these digital dataset. 
Even when they are present, they will not generate data of the same \textit{utility} and often be considered outliers\cite{schlosser2021biases}.
In these situations we need to pay attention to what is important, not just what is quantifiable\cite{merry2015quantification,thomas2020problem}.
Unless these data inequalities and biases are actively taken into account, large populations will be excluded from future analyses, independent of which metrics are applied to measure algorithmic performance.

\section*{The way forward}
Computational tools from machine learning to network science can bring tremendous potential to the humanitarian and development sectors, however, they also bring a lot of unknowns.
For these new methods to be routinely applied in development programs, they need to be put through rigorous evaluation processes.
In order to minimize any potential harms, testing and evaluation needs to be done prior to releasing new models.
The old Silicon Valley mantra of \textit{"moving fast and breaking things"} does not work for international development, where decisions directly affect the life and well-being of populations.
As such, it is vital to ensure that new methodologies produce replicable, explainable, transparent, and generalizable results, and that potential limitations, biases, and shortcomings are uncovered and documented.
This includes improving both scientific and ethical practices within AI and ML\cite{raji2020closing}.

Digital data-sources offer the opportunity to discover insights at unprecedented scale and speed.
This can lead to a lot of good.
For instance, data-poor areas can now effortlessly be mapped.
However, it is critical to recognize and scrutinize how digital data are imperfect and potentially biased.
For example, in certain contexts mobile phones are predominantly owned and used by men\cite{blumenstock2012divided}.
Does this mean that insights derived from such data will mainly represent male demographics? How will deprivations suffered by women and children be represented by models based on this data?
Furthermore, what will happen once gender equality is achieved, will the earlier trained sophisticated algorithms break down and return incorrect and biased estimates?
Only rigorous scientific studies where the representativity of the data is inspected can answer these questions.
Studying these systems of multiple interactive components through a complex systems lens could produce novel insights.

Standardized data formats and open repositories that contain machine readable development data would empower practitioners and researchers working on addressing these issues.
In addition, standardized ways of sharing AI models, that go beyond posting code on GitHub, would benefit future replication, transferability, and robustness studies. 

This is not an impossible task, but it requires careful planing, auditable standards\cite{coppi2021explicability}, long-term partnerships, transparent guidelines, in-house AI expertise, and an understanding of the risks and pitfalls of data-driven technologies.
For instance, Eurostat, the statistical office of the European Union, is exploring new datasources for official statistics\cite{eurostat} by comparing results extracted from big-data to statistical gold-standards, while at the same time being mindful of sampling biases, the volatility of big data, and that a high volume of data does not necessary guarantee high data quality.

Overall, the diffusion of data science and AI techniques into the realm of international development constitutes a unique opportunity to bring powerful new techniques to the fight against inequalities and vulnerabilities.
However, improving living conditions and creating lasting change can only be accomplished through closer collaborations between academic, private, governmental, and non-governmental actors.
Local stakeholders need to be engaged, trust needs to be built, and local data-science ecosystems need to be strengthened to ensure that proposed solutions are adapted to local contexts\cite{abebe2021narratives}.
There are no one-size-fits-all AI solutions, so we need to know what to implement where, and which solutions should never be built\cite{birhane2020algorithmic}. 

The biggest impediment to this is the capacity gap.
AI talent is far removed from humanitarian and development organizations, and limited funding is available to increase that capacity. 
Similarly, humanitarian and social science expertise is often limited, and in some cases entirely missing, from AI hubs.
Many of the above problems are a consequence of the distance
between these communities, and of the misalignment of their incentives. 
This gap needs to be bridged from both sides, more resources need to be put towards bringing AI expertise into the humanitarian domain, and vice versa in infusing humanitarian thinking into AI\cite{walther2021technology}.
Today, it is clear that we need to devise data-driven methodologies, which are fair and capable of producing actionable insights, while protecting all fundamental human rights\cite{un1948}, including the right to privacy and the right to non-discrimination.
The price of innovating should not come at the cost of eroding these rights. 

\linespread{1}
\section*{References}
\small
\bibliography{references}
\bibliographystyle{naturemag}


\begin{addendum}
\item M.C. was supported by the Ministry of Universities of the Government of Spain under the program 'Convocatoria de Ayudas para la recualificacion del sistema universitario español para 2021-2023' from the Universidad Carlos III de Madrid, dated July 1, 2021.
E.M. would like to thank Alex 'Sandy' Pentland for helpful discussions and comments. E.M. acknowledges support by Ministerio de Ciencia e Innovaci\'on/Agencia Espa\~nola de Investigaci\'on   (MCIN/AEI/10.13039/501100011033) through grant PID2019-106811GB-C32 the National Science Foundation under Grant No.~2218748.
 \item[Correspondence] Correspondence and requests for materials
should be addressed to V.S.~(email: vedransekara@gmail.com) and M.G-H.~(email: mherranz@unicef.org).
\end{addendum}


\end{document}